\documentclass[aps,twocolumn,groupedaddress,nofootinbib,floatfix]{revtex4-1}
\usepackage{graphicx}
\usepackage{color}
\usepackage{pifont}
\usepackage{amsmath,amsfonts,amssymb}
\usepackage{bbold}

\newcommand{\met}{\mbox{$\protect \raisebox{.3ex}{$\not$}E_T$}}

\newcommand{\smo}{{\textstyle \scriptscriptstyle -}1}
\newcommand{\smt}{{\textstyle \scriptscriptstyle -}2}

\newcommand{\smpr}{{\scriptscriptstyle(')}}

\DeclareMathOperator{\tr}{Tr}

\begin{document}

\title{Laboratory-frame tests of quantum entanglement in $H \to WW$}

\author{J. A. Aguilar-Saavedra}
\affiliation{Instituto de F\'\i sica Te\'orica, IFT-UAM/CSIC, c/ Nicolás Cabrera 13-15, 28049 Madrid}

\begin{abstract}
Quantum entanglement between the two $W$ bosons resulting from the decay of a Higgs boson may be investigated in the dilepton channel $H \to WW \to \ell \nu \ell \nu$ using laboratory-frame observables that only involve the charged leptons $\ell=e,\mu$. The dilepton invariant mass distribution, already measured by the ATLAS and CMS Collaborations at the LHC, can be used to observe the quantum entanglement of the $WW$ pair with a statistical sensitivity of $7\sigma$ with Run 2 data, and of $6\sigma$ when including theoretical systematics. As a by-product, the relation between $W$ rest frame (four-dimensional) angular distributions, $H \to WW$ decay amplitudes, and spin correlation coefficients, is written down.
\end{abstract}

\maketitle

\section{Introduction}

Ten years after the discovery of the Higgs boson by the ATLAS and CMS experiments~\cite{ATLAS:2012yve,CMS:2012qbp}, the statistics collected at the Large Hadron Collider (LHC) allows to test its properties in many production and decay modes~\cite{ATLAS:2022vkf,CMS:2022dwd}. The main goal is to determine from experimental data whether the 125 GeV particle discovered corresponds to the Standard Model (SM) Higgs or not; in particular, whether it is the first discovered particle of an extended scalar sector. In addition, tests of the quantum properties of its decay, i.e. the quantum entanglement and possible violation of Bell inequalities~\cite{Bell:1964kc} are recently attracting attention~\cite{Barr:2021zcp,Aguilar-Saavedra:2022wam}. While there is no experimental evidence to call into question the validity of quantum mechanics, testing it at the energy frontier is of high relevance. And the proposed tests often yield, as a by-product, new observables that might also be useful in searches for physics beyond the SM. Several studies in this regard have been performed for the entangled state of a top quark-antiquark pair~\cite{Afik:2020onf, Fabbrichesi:2021npl, Severi:2021cnj, Aoude:2022imd,Afik:2022kwm,Aguilar-Saavedra:2022uye,Fabbrichesi:2022ovb,Afik:2022dgh}. 

The decays $H \to VV$, $V = W,Z$ (with one of the weak bosons off shell) provide the ideal environment to test Higgs properties. In particular, quantum entanglement leaves its imprint in the spin correlation between the daughter weak bosons: because the Higgs is a spin-zero particle, the $VV$ pair is produced in a state of vanishing total angular momentum. If the $V$ were produced at rest in the Higgs rest frame, the orbital angular momentum would also vanish and the $VV$ pair would be in a maximally-entangled spin-singlet state. In practice, the two bosons are produced quite close to a spin singlet.

The possible violation of Bell-like inequalities in $H \to W^+W^-$ has been addressed in Ref.~\cite{Barr:2021zcp}, focusing on the dilepton final state $W^+W^- \to \ell^+ \nu \ell^- \nu$, $\ell = e,\mu$, and using for spin measurements $W$-rest frame angular distributions. This implicitly assumes that the $W$ rest frames can be determined, which is not obvious because the two neutrinos are undetected, and only the sum of their transverse momenta can be identified with the missing transverse energy (MET) in the event.\footnote{In top pair production in the dilepton final state, $t \bar t \to W^+ b W^- \bar b \to \ell^+ \nu b \ell^- \nu b$, the kinematics can be fully reconstructed, up to discrete ambiguities, because there are six unknowns (the three-momenta of the two neutrinos) and four constraints (the invariant masses of $t$, $\bar t$, $W^+$ and $W^-$, plus the two MET constraints). In $H \to W^+ W^- \to \ell^+ \nu \ell^- \nu$ there still are six unknowns but only four constraints, two from the MET and two from the masses of $H$ and the on-shell $W$ boson.} A reconstruction of the $W$ momenta using a kinematical fit or a multivariate method, e.g. a neural network, faces the difficulty of selecting a `best solution' for the neutrino momenta within a two-dimensional manifold of possible solutions allowed by the kinematical constraints. The procedure adopted for the $W$ momenta determination might wash out the information from their spin that is transferred to the daughter leptons, but it may be worth exploring this kind of methods.

On the other hand, in this paper we propose tests of the $WW$ entanglement based on laboratory-frame observables, such as (i) the dilepton invariant mass $m_{\ell \ell}$; (ii) the angular separation between the leptons in the plane orthogonal to the beam axis $\phi_{\ell \ell}$; (iii) their pseudo-rapidity difference $\eta_{\ell \ell}$. We note that a similar approach was followed to establish the existence of spin correlations in $t \bar t$ production at the LHC. In the dilepton decay $t \bar t \to \ell^+ \nu b \ell^- \nu b$, the azimuthal angle difference between the charged leptons in the laboratory frame was identified in Ref.~\cite{Mahlon:2010gw} to be quite sensitive to discriminate the SM versus the no-correlation scenario. Subsequently, this distribution was measured by the ATLAS~\cite{ATLAS:2012ao} and CMS~\cite{CMS:2013roq} Collaborations to establish the existence of spin correlations.

In order to investigate the feasibility of the entanglement measurement, in section~\ref{sec:2} we use the helicity amplitude formalism of Jacob and Wick~\cite{Jacob:1959at} to write down the general prediction for polarisation and spin correlations in $H \to VV$, in terms of the decay amplitudes. In section~\ref{sec:3} we discuss the  $m_{\ell \ell}$, $\phi_{\ell \ell}$ and $\eta_{\ell \ell}$ distributions as a test of the $W^+ W^-$ entanglement. The experimental prospects to disentangle the two options are examined in section~\ref{sec:4}, and our results are discussed in section~\ref{sec:5}.

\section{$H \to VV$ and angular momentum}
\label{sec:2}

In order to be more general, let us consider the decay $H \to V_1 V_2$, with $V_1 V_2 = ZZ, W^+W^-$ and label as $f = \ell,\nu$ the decay products of the weak bosons.\footnote{In this section we label the two bosons with subindices $1,2$ to emphasise that we consider them as distinguishable. Even when both are $Z$ bosons, one of them is quite close to its mass shell while the other one is well below.} Using the helicity amplitude formalism~\cite{Jacob:1959at} we can write the amplitudes for the decay $H \to V_1 V_2 \to f_1 f_1' f_2 f_2'$ as
\begin{eqnarray}
A_{\lambda_1 \lambda_1' \lambda_2 \lambda_2'} & = & \sum_{\Lambda_1 \Lambda_2} a_{\Lambda_1 \Lambda_2} b_{\lambda_1 \lambda_1'} c_{\lambda_2 \lambda_2'} \notag \\
& & \times D_{\Lambda_1 \lambda}^{1*} (\phi_1,\theta_1,0) D_{\Lambda_2 \lambda'}^{1*} (\bar \phi_2,\bar \theta_2,0)
\label{ec:AH}
\end{eqnarray}
where $\Lambda_{1,2}$ are the helicities of $V_1$ and $V_2$, respectively, with $\Lambda_1 = \Lambda_2$ by angular momentum conservation; $\lambda_i^\smpr$ are the helicities of $f_i^\smpr$, and $\lambda^\smpr = \lambda_1^\smpr - \lambda_2^\smpr$. 
Note that the off-shell $V$ propagator includes a `scalar' component that produces distinct terms in the angular distributions. However, when coupled to massless external fermions the scalar component vanishes; therefore, we can safely consider the off-shell $W$ as a spin-1 particle~\cite{Groote:2012jq}. In the above equation, the angular dependence is given by the well-known Wigner functions~\cite{wigner}
\begin{equation}
D^j_{m'm}(\alpha,\beta,\gamma) \equiv \langle jm' | e^{-i \alpha J_z} e^{-i \beta J_y} e^{-i \gamma J_z} | jm \rangle \,,
\end{equation}
and  $a_{\Lambda_1 \Lambda_2}$, $b_{\lambda_1 \lambda_1'}$ and $c_{\lambda_2 \lambda_2'}$ are constants that depend on the helicity combination considered. For $Z$ bosons there are two non-zero combinations $(\lambda_i,\lambda_i') = (\pm 1/2,\mp 1/2)$, and the corresponding $b$ and $c$ constants are related by the ratio of the left- and right-handed couplings to leptons, $g_R^\ell \,:\, g_L^\ell$. For $W$ bosons there is only one such combination because the coupling is purely left-handed.
The angles $(\theta_1,\phi_1)$ are the polar coordinates of the three-momentum of $f_1$ in the $V_1$ rest frame, and likewise the angles $(\bar \theta_2, \bar \phi_2)$ are the polar coordinates of the three-momentum of $f_2$ in the $V_2$ rest frame. Using 
\begin{equation}
D^j_{m'm}(\alpha,\beta,\gamma) = e^{-i \alpha m'} e^{-i \gamma m} d^j_{m'm}(\beta)
\end{equation}
the amplitudes can be simplified to 
\begin{eqnarray}
A_{\lambda_1 \lambda_1' \lambda_2 \lambda_2'} & = & \sum_{\Lambda_1 \Lambda_2} a_{\Lambda_1 \Lambda_2} b_{\lambda_1 \lambda_1'} c_{\lambda_2 \lambda_2'}  \notag \\
& & \times e^{i \Lambda_1(\phi_1 + \bar \phi_2)} d_{\Lambda_1 \lambda}^{1} (\theta_1) d_{\Lambda_2 \lambda'}^{1} (\bar \theta_2) \,.
\label{ec:AH2}
\end{eqnarray}
The orientation of the three axes in the reference systems for the $V_{1,2}$ rest frames stems from the precise way in the helicity states are defined (see for example~\cite{libro}). If we set a reference system $(x,y,z)$ in the Higgs rest frame, in which the $V_1$ boson three-momentum has angular coordinates $(\theta,\phi)$,
the reference system $(x_1,y_1,z_1)$ in the $V_1$ rest frame has the axes as follows:
\begin{itemize}
\item The $\hat z_1$ axis is in the direction of the $V_1$ boson three-momentum in the Higgs rest frame, $\hat z_1 = \sin \theta \cos \phi \, \hat x + \sin \theta \sin \phi \, \hat y + \cos \theta \, \hat z$.
\item The $\hat y_1$ axis is in the $xy$ plane, making an angle $\phi$ with the $\hat y$ axis: $\hat y_1 = - \sin \phi \, \hat x + \cos \phi \, \hat y$.
\item The $\hat x_1$ axis is orthogonal to both, $\hat x_1 = \hat y_1 \times \hat z_1 = \cos \theta \cos \phi \, \hat x + \cos \theta \sin \phi \, \hat y - \sin \theta \hat z$.  
\end{itemize}
For the reference system $(x_2,y_2,z_2)$ in the $V_2$ rest frame one has a similar definition. However, for entanglement studies it is convenient to use {\it the same} definition of axes in both rest frames~\cite{Aguilar-Saavedra:2022wam}. A simple computation shows that 
\begin{equation}
\hat x_2 = \hat x_1 \,,\quad \hat y_2 = - \hat y_1 \,,\quad \hat z_2 = - \hat z_1 \,, 
\end{equation}
therefore the polar coordinates of the $f_2$ three-momentum in the $V_2$ rest frame, with respect to the axes defined for the $V_1$ rest frame, are
\begin{equation}
\theta_2 = \pi - \bar \theta_2 \,, \quad \phi_2 = - \bar \phi_2 \,.
\end{equation}
Using the symmetry properties of the $d_{m'm}^j$ functions, the amplitudes (\ref{ec:AH2}) can be rewritten as
\begin{eqnarray}
A_{\lambda_1 \lambda_1' \lambda_2 \lambda_2'} & = & \sum_{\Lambda_1 \Lambda_2} a_{\Lambda_1 \Lambda_2} b_{\lambda_1 \lambda_1'} c_{\lambda_2 \lambda_2'}  \notag \\
& & \times e^{i \Lambda_1(\phi_1 - \phi_2)} d_{\Lambda_1 \lambda}^{1} (\theta_1) d_{-\Lambda_2 \lambda'}^{1} (\theta_2) \,.
\label{ec:AH3}
\end{eqnarray}
The differential cross section is proportional to the squared amplitude summed over final state helicities,
\begin{equation}
\frac{d\sigma}{d\Omega_1 d\Omega_2} \propto 
\sum_{\lambda_1 \lambda_1' \lambda_2 \lambda_2'} |A_{\lambda_1 \lambda_1' \lambda_2 \lambda_2'}|^2 \,,
\label{ec:dist0}
\end{equation}
with $d\Omega_i = d\cos \theta_i d\phi_i$.
Rather than writing the full expression, it is convenient to match it to a general parameterisation for the decay 
$V_1 V_2 \to f_1 f_1' f_2 f_2'$ in terms of polarisation and spin correlation coefficients.

A convenient parameterisation of the spin density operator for the $V_1 V_2$ pair can be found in terms of the identity and 8 irreducible tensor operators $T^L_M$, with $L=1,2$ and $-L \leq M \leq L$, acting on the three-dimensional spin space for each boson~\cite{Aguilar-Saavedra:2022wam}. For convenience we normalise $T^L_M$ such that $\tr\left[T^L_M \left(T^L_M\right)^{\dagger}\right] = 3 $, where $\left(T^L_M\right)^{\dagger}=(-1)^M \, T^L_M $ (note the change of normalisation with respect to Refs.~\cite{Aguilar-Saavedra:2015yza,Aguilar-Saavedra:2017zkn}). Specifically, the $T^L_M$ operators are defined as
\begin{align}
& T^1_{\pm 1} = \mp \frac{\sqrt{3}}{2}  (S_1 \pm i S_2) \,,\quad T^1_0 = \sqrt{\frac{3}{2}} S_3 \,, \notag \\
& T^2_{\pm 2} = \frac{2}{\sqrt{3}} (T_{\pm 1}^1)^2 \,, \quad
    T^2_{\pm 1} = \sqrt{\frac{2}{3}} \left[T_{\pm 1}^1 T_{0}^1 + T_{0}^1 T_{\pm 1}^1 \right] \,, \notag \\
& T^2_0 = \frac{\sqrt{2}}{3} \left[T_1^1 T_{-1}^1 + T_{-1}^1 T_{1}^1 + 2 (T_{0}^1)^2 \right] \,,
\end{align}
with $S_i$ the usual spin operators. In terms of these, the spin density operator reads~\cite{Aguilar-Saavedra:2022wam}
\begin{eqnarray}
\rho & = & \frac{1}{9}\left[
\mathbb{1}_3 \otimes \mathbb{1}_3 + A^1_{LM} \, T^L_{M} \otimes \mathbb{1}_3 + A^2_{LM} \, \mathbb{1}_3 \otimes T^L_{M} \right. \notag \\
& &  \left. + C_{L_1 M_1 L_2 M_2} \, T^{L_1}_{M_1} \otimes T^{L_2}_{M_2}
\right] \,,
\label{rhoAC}
\end{eqnarray}
where an implicit sum over all indices is understood. The coefficients satisfy the relations
\begin{align}
& (A^{1,2}_{LM})^* = (-1)^M \, A^{1,2}_{L \; -M} \,, \notag \\
& (C_{L_1 M_1 L_2 M_2})^* = (-1)^{M_1+M_2} \, C_{L_1 \; -M_1 L_2 \; -M_2} \,.
\end{align}
The angular distribution corresponding to this density operator can be compactly written~\cite{Aguilar-Saavedra:2022wam} in terms of spherical harmonics $Y_\ell^m$,
\begin{align}
& \frac{1}{\sigma}\frac{d\sigma}{d\Omega_1d\Omega_2} = \frac{1}{(4\pi)^2}\left[ 1 +A_{LM}^1 B_L Y_L^M(\theta_1, \phi_1) \right. \notag \\
& \quad + A_{LM}^2 B_L Y_L^M(\theta_2, \phi_2)  \notag \\
& \quad \left. + C_{L_1M_1L_2M_2} B_{L_1}B_{L_2} Y_{L_1}^{M_1}(\theta_1, \phi_1)Y_{L_2}^{M_2}(\theta_2, \phi_2)  \right] \,,
\label{ec:dist4D}
\end{align}
with $B_1$, $B_2$ constants. For $V_1 = V_2 = Z$, and taking $f_{1,2}$ as the negative leptons, one has
\begin{eqnarray}
B_1=-\sqrt{2\pi} \eta_\ell\ , \ \ \ B_2=\sqrt{\frac{2\pi}{5}}
\label{ec:BL}
\end{eqnarray}
with~\cite{Aguilar-Saavedra:2017zkn}
\begin{equation}
\eta_\ell = \frac{(g_L^\ell)^2 - (g_R^\ell)^2}{(g_L^\ell)^2 + (g_R^\ell)^2} = \frac{1-4 s_W^2}{1-4 s_W^2 + 8 s_W^4} \simeq 0.13 \,,
\label{ec:etal}
\end{equation}
$s_W$ being the sine of the weak mixing angle. For $V_2=W^-$, and taking $f_2$ as the negative lepton, $B_{1,2}$ are as in (\ref{ec:BL}) setting $\eta_\ell = 1$. For $V_1 = W^+$, and taking $f_1$ as the positive lepton (which is an anti-fermion), $B_{1,2}$ are as in (\ref{ec:BL}) but setting instead $\eta_\ell = -1$. Thus the parameterisation (\ref{ec:dist4D}) with the above conventions summarises the double distribution for both $ZZ$ and $W^+ W^-$ in terms of polarisation and spin correlation coefficients.

One can match the expression (\ref{ec:dist4D}) to (\ref{ec:dist0}) properly normalised, to identify the non-zero coefficients. Let us define
\begin{equation}
\mathcal{N} = |a_{1 1}|^2 + |a_{00}|^2 + |a_{\smo \smo}|^2 \,.
\end{equation}
Within the SM, CP is conserved in the $H \to V_1 V_2$ decay at the leading order (LO), and only 
\begin{align}
& A_{20}^1 = A_{20}^2 = \frac{1}{\sqrt 2} \frac{1}{\mathcal{N}} \left[ 
|a_{1 1}|^2 + |a_{\smo \smo}|^2 - 2 |a_{00}|^2 \right] \,, \notag \\
& C_{1010} = - \frac{3}{2} \frac{1}{\mathcal{N}} \left[ 
|a_{1 1}|^2 + |a_{\smo \smo}|^2 \right] \,, \notag \\
& C_{2020} = \frac{1}{2} \frac{1}{\mathcal{N}} \left[ 
|a_{1 1}|^2 + |a_{\smo \smo}|^2 + 4 |a_{00}|^2 \right] \,, \notag \\
& C_{2 2 2 \smt} = C_{2 \smt 22}^* = 3 \frac{1}{\mathcal{N}} \; a_{11} a_{\smo \smo}^* \,, \notag \\
& C_{111 \smo} = - C_{212 \smo} = C_{1 \smo 11}^* = - C_{2 \smo 21}^* \notag \\
& \quad = - \frac{3}{2} \frac{1}{\mathcal{N}} \left[ 
a_{11} a_{00}^* + a_{00} a_{\smo \smo}^* \right] 
\label{ec:C}
\end{align}
are non-vanishing.
CP-violating effects in the SM arise beyond the LO but are at the level of $10^{-5}$~\cite{Gritsan:2022php}, so CP conservation is an excellent approximation.
If CP is broken in the $H \to V_1 V_2$ decay due to effects beyond the SM, additional terms appear
\begin{align}
& A_{10}^1 = - A_{10}^2 = \sqrt{\frac{3}{2}} \frac{1}{\mathcal{N}} \left[ 
|a_{1 1}|^2 - |a_{\smo \smo}|^2 \right] \,, \notag \\
& C_{1020} = - C_{2010} = \frac{\sqrt 3}{2}  \frac{1}{\mathcal{N}} \left[ 
|a_{1 1}|^2 - |a_{\smo \smo}|^2 \right] \,, \notag \\
& C_{1 \smo 2 1} = - C_{2 \smo 11} = C_{112 \smo}^* = - C_{211 \smo}^* \notag \\
& \quad = \frac{3}{2} \frac{1}{\mathcal{N}} \left[ 
a_{00} a_{11}^* - a_{\smo \smo} a_{00}^* \right] \,.
\end{align}

\section{Entanglement in the laboratory frame}
\label{sec:3}

\begin{figure}[t!]
\begin{center}
\begin{tabular}{c}
\includegraphics[width=8cm,clip=]{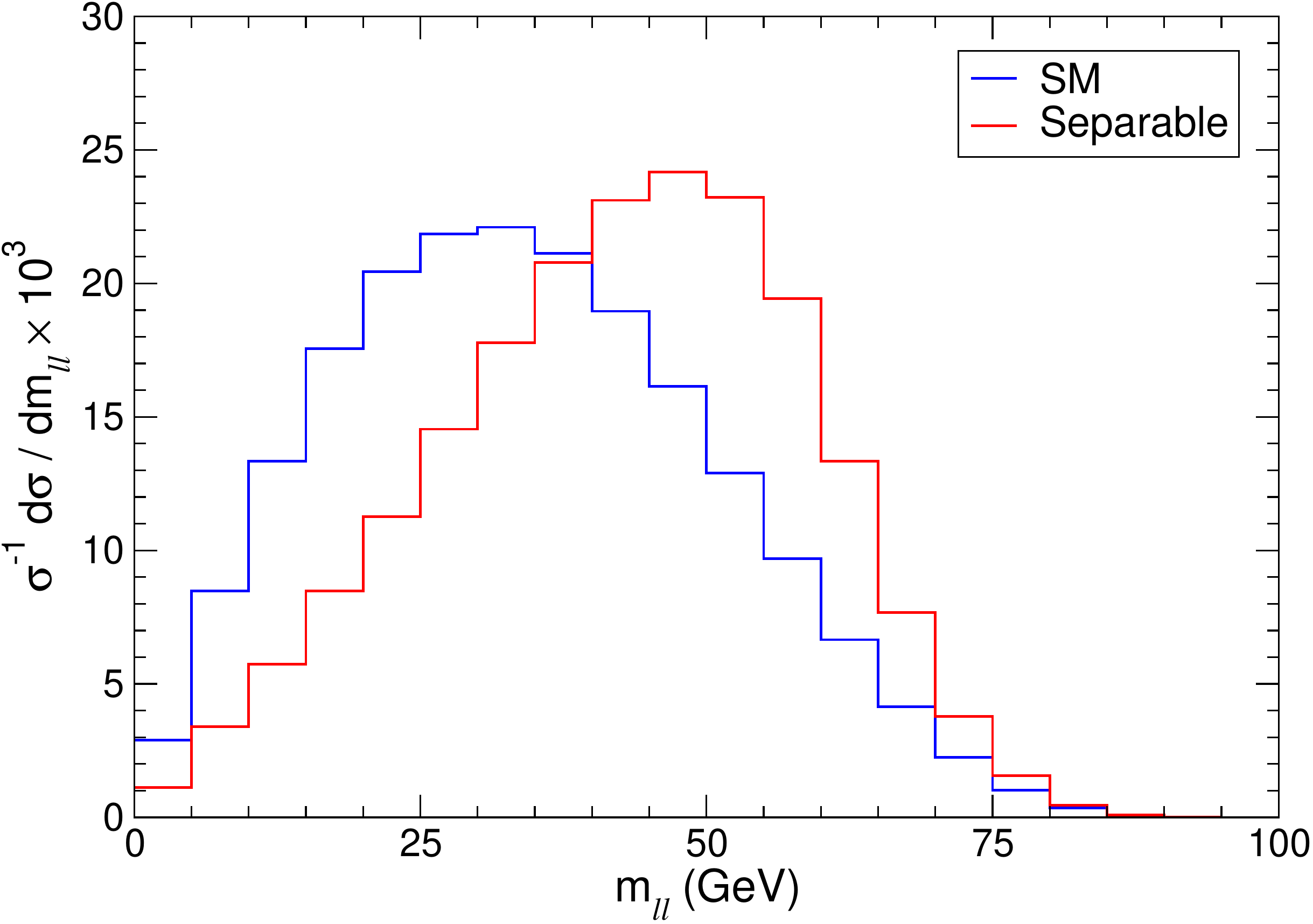} \\
\includegraphics[width=8cm,clip=]{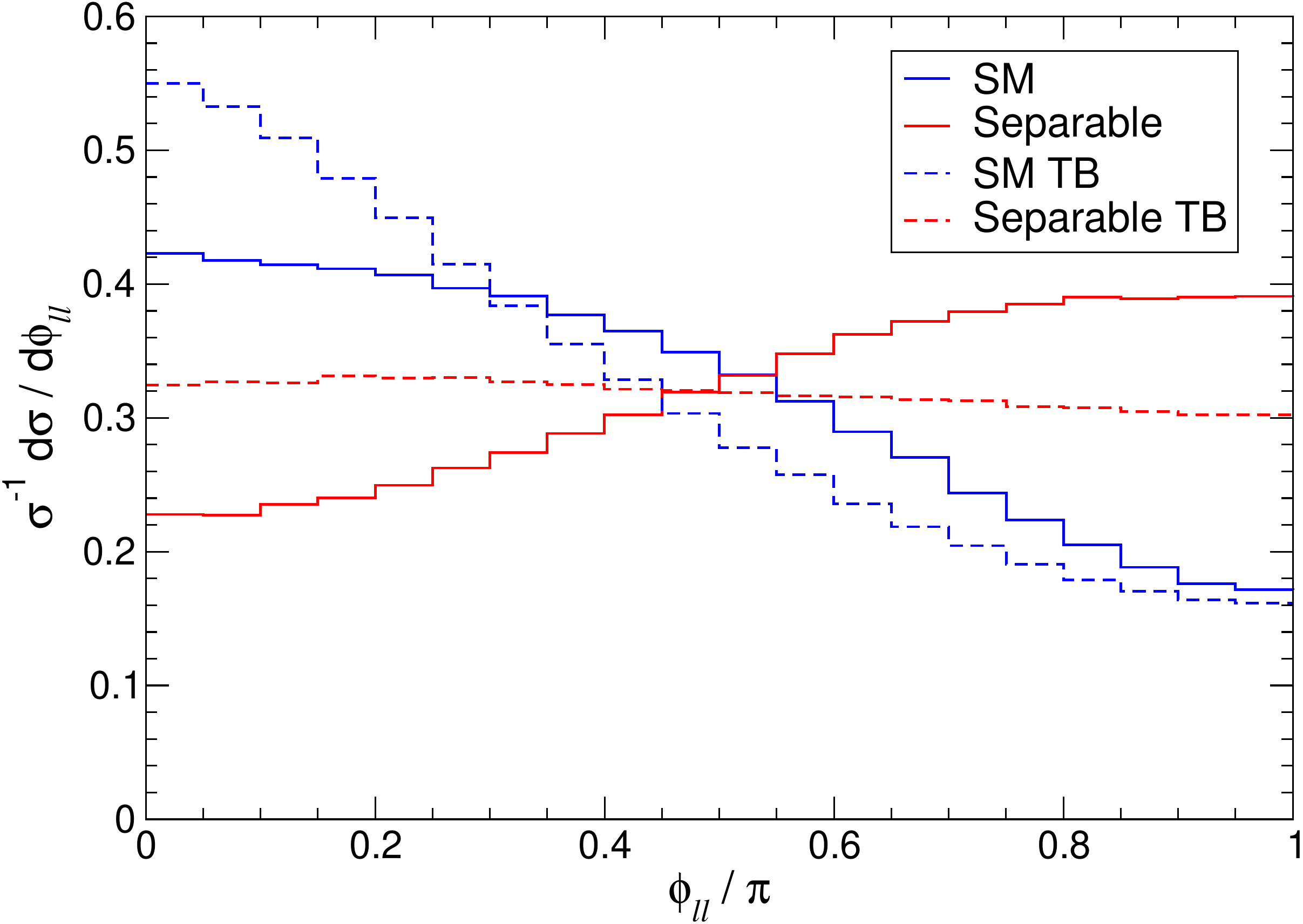}  \\
\includegraphics[width=8cm,clip=]{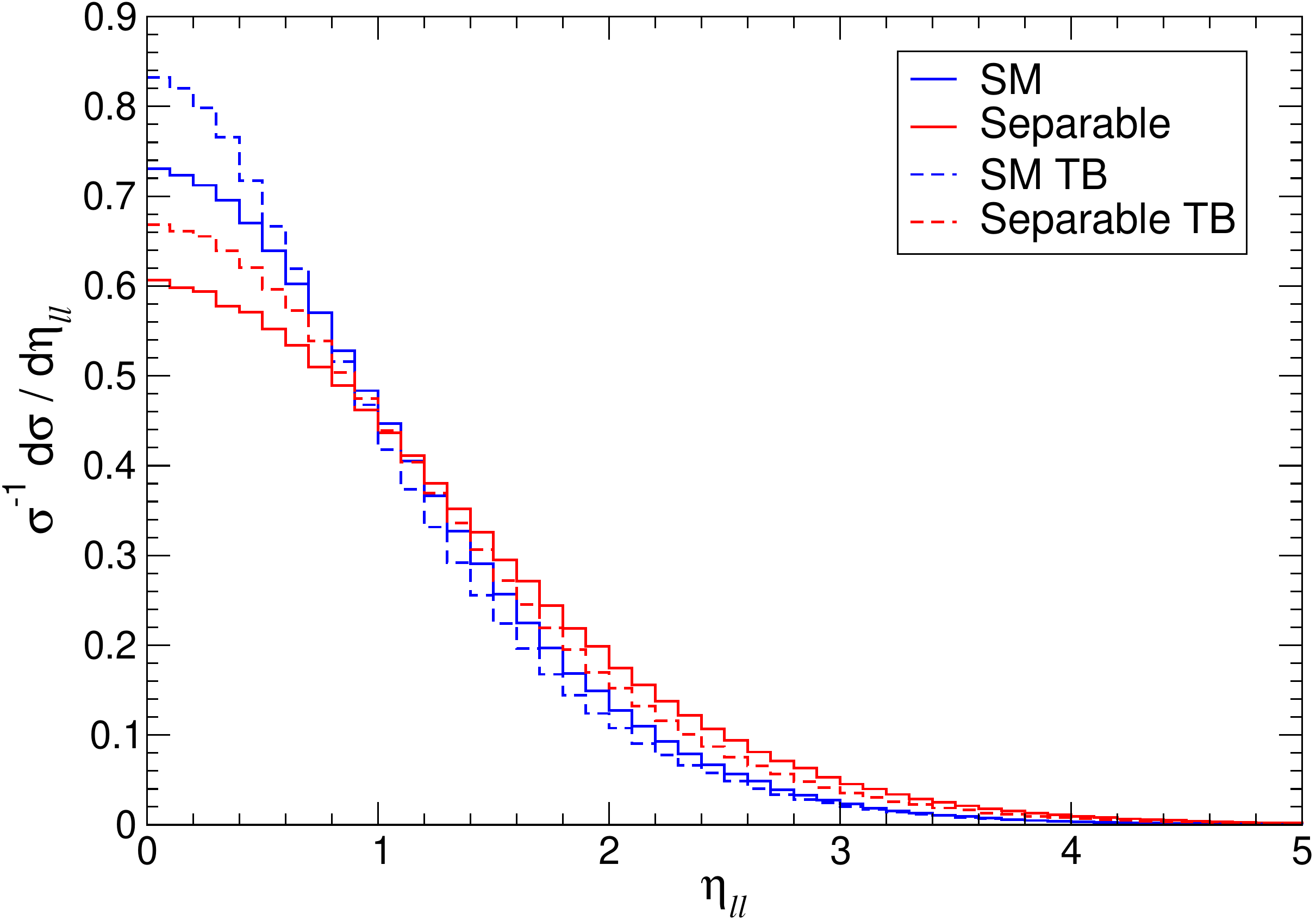}
\end{tabular}
\caption{Dilepton observables:  invariant mass (top), azimuthal angle difference (middle) and pseudo-rapidity difference (bottom). The dashed lines labeled as `TB' show the distributions with $p_T^H = 20$ GeV.}
\label{fig:distP}
\end{center}
\end{figure}

From now on we focus on the decay $H \to W^+ W^-$. 
Reference~\cite{Aguilar-Saavedra:2022wam} showed that a necessary and sufficient condition for entanglement (which applies too to this decay) is $C_{222 \smt} \neq 0$. Since $a_{11} = a_{\smo \smo}$, Eq.~(\ref{ec:C}) implies that the $W^+ W^-$ pair is entangled as long as these amplitudes are non-vanishing. The test of entanglement then consists in comparing between (i) the SM; (ii) a decay where the $W$ bosons have longitudinal polarisation, $a_{11} = a_{\smo \smo} = 0$, $a_{00} \neq 0$.

We generate a sample of $H \to W^+ W^- \to e^\pm \nu \mu^\mp \nu$ at the LO in the SM using {\scshape MadGraph}~\cite{Alwall:2014hca}, with the implementation of the $gg \to H$ loop as a contact interaction. A second sample is obtained from the former by modifying both $W$ decays following the CAR method~\cite{Aguilar-Saavedra:2022kgy} so that the angular distributions of the decay products correspond to zero helicity. This modification of the decay gives the exact result for polarised $H \to W^+ W^-$ decays, because the kinematics of the $W W$ pair in the Higgs rest frame is independent of the polarisation.

For the two cases (SM versus longitudinal $W$ polarisations) Fig.~\ref{fig:distP} shows the distributions of the three variables of interest: (i) the dilepton invariant mass; (ii) the angle $\phi_{\ell \ell}$ between the two leptons in the plane transverse to the beam axis; (iii) the modulus of the pseudo-rapidity difference $\eta_{\ell \ell} = |\eta_{\ell^+} - \eta_{\ell^-}|$. The difference in the $\phi_{\ell \ell}$ distributions between the two possibilities is striking and is caused by the two charged leptons being preferably emitted in the same direction when the $W^+ W^-$ pair has like helicities. However, this variable is quite sensitive to boosts in the transverse direction due to initial state radiation, which causes the Higgs boson to be produced with non-zero transverse momentum $p_T^H$. We illustrate this effect by applying a boost in the transverse plane that gives $p_T^H = 20$ GeV. The resulting distributions are shown in dashed lines. The modifications in $\eta_{\ell \ell}$ and especially $\phi_{\ell \ell}$ are important but one can see that the differences between the SM and the separable case are maintained at this level.
Note that $m_{\ell \ell}$ is Lorentz invariant and therefore is unaffected by non-zero $p_T^H$, and it can also be measured in the laboratory frame.

\section{Sensitivity to entanglement}
\label{sec:4}

In contrast to the $H \to ZZ$ decay mode studied in Ref.~\cite{Aguilar-Saavedra:2022wam}, the background for $H \to WW$ is much larger than this signal. This has two consequences that may jeopardise the observation of entanglement. First, the kinematical selection necessary to suppress the background may shape the $H$ signal and dilute the differences between the SM and separable case. Second, the statistical uncertainties in the background, larger than the signal, make it harder (and statistically less significant) the discrimination between the two hypotheses. 

For the sensitivity estimation we restrict ourselves to the different-flavour final state, $H \to WW \to e^\pm \nu \mu^\mp \nu$, for which the background is much smaller, and dominated by electroweak $WW$ production when both charged leptons are energetic~\cite{CMS:2022uhn}. The  $H \to WW \to e^\pm \nu \mu^\mp \nu$ processes are generated as described in the previous section, with a Monte Carlo statistics of $2 \times 10^6$ events. For the background $pp \to WW \to e^\pm \nu \mu^\mp \nu$  we also use {\scshape MadGraph} at the LO, with a Monte Carlo statistics of $3 \times 10^6$ events. The parton-level event samples are showered and hadronised with {\scshape Pythia} 8.3~\cite{Sjostrand:2007gs} and a fast detector simulation is performed with {\scshape Delphes}~\cite{deFavereau:2013fsa}, using the default CMS card. 

Despite the Monte Carlo generation is done at the LO, we use higher-order predictions of the cross sections to calculate the expected number of events. The Higgs cross section in gluon-gluon fusion at next-to-next-to-next-to-leading order is 48.61 pb at a centre-of-mass energy of 13 TeV~\cite{Cepeda:2019klc}, and the Higgs branching ratio decay into $e^\pm \nu \mu^\mp \nu$ is $5.038 \times 10^{-3}$~\cite{Cepeda:2019klc}, yielding an overall cross section times branching ratio of 245 fb at 13 TeV. The $WW$ cross section is normalised to next-to-next-to-leading order (NNLO) with a $K$-factor of 1.4~\cite{Grazzini:2019jkl}, and for the final state considered is 2.6 pb.

We follow Ref.~\cite{CMS:2022uhn} to implement a kinematical selection on charged leptons:
\begin{itemize}
\item Both leptons must have pseudo-rapidities $|\eta| \leq 2.5$, the leading one with transverse momentum $p_T \geq 25$ GeV and the sub-leading one with $p_T \geq 20$ GeV.
\item The transverse momentum and invariant mass of the dilepton pair must be above the minimum thresholds $p_T^{\ell \ell} \geq 30$ GeV, $m_{\ell \ell} \geq 12$ GeV, respectively, and the missing energy $\met \geq 20$ GeV.
\item The transverse mass of the event (with the usual definition) must be $m_T \geq 60$ GeV; the transverse mass constructed using only the sub-leading lepton (see Ref.~\cite{CMS:2022uhn} for the precise definition) is $m_{T2} \geq 30$ GeV.
\end{itemize}
With this selection, the electroweak $WW$ background is dominant~\cite{CMS:2022uhn}; therefore, we can safely ignore the rest of them, mainly $t \bar t$ and $tW$, to obtain a realistic estimate of the statistical sensitivity to quantum entanglement.

\begin{figure}[t]
\begin{center}
\begin{tabular}{c}
\includegraphics[width=8cm,clip=]{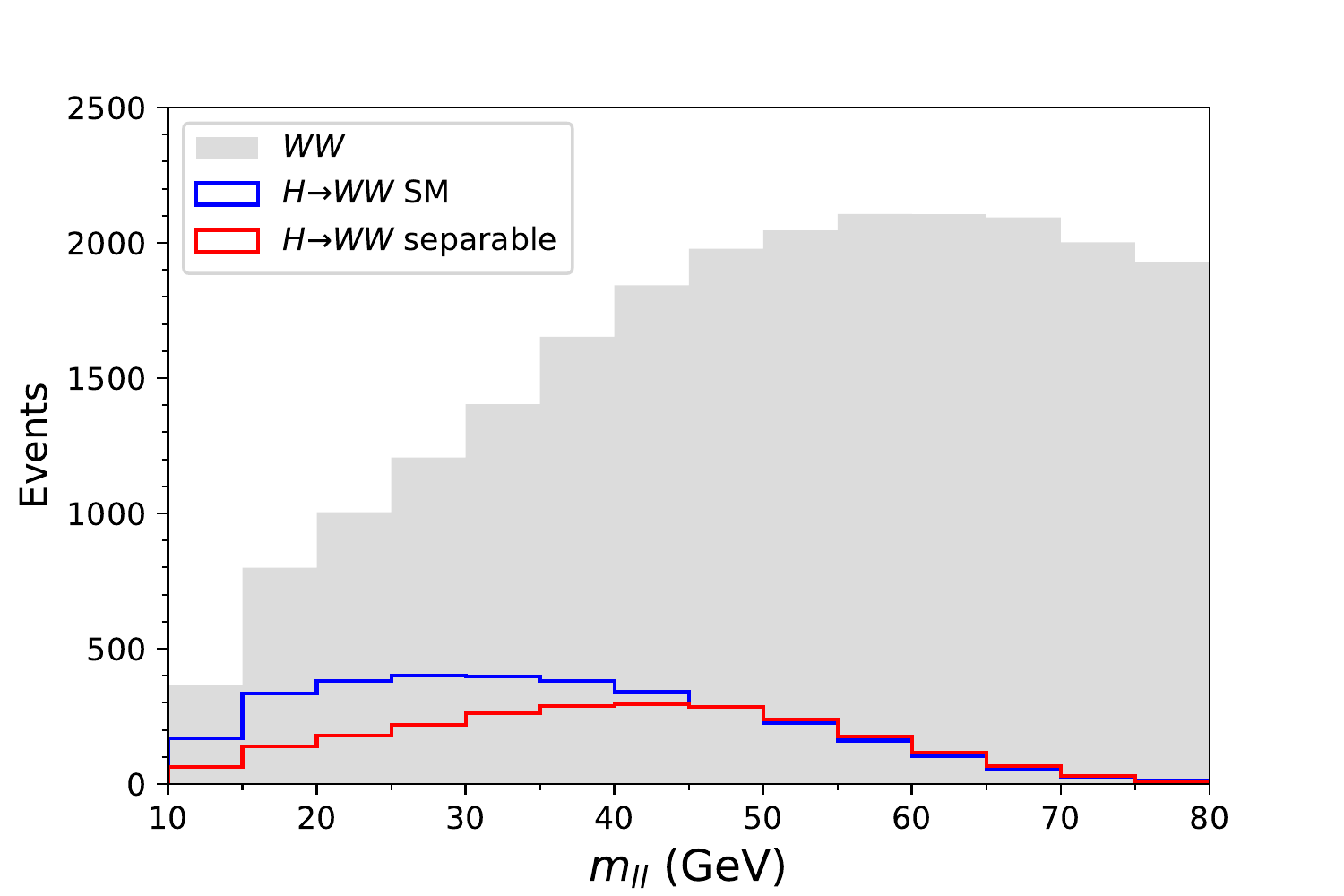}  \\
\includegraphics[width=8cm,clip=]{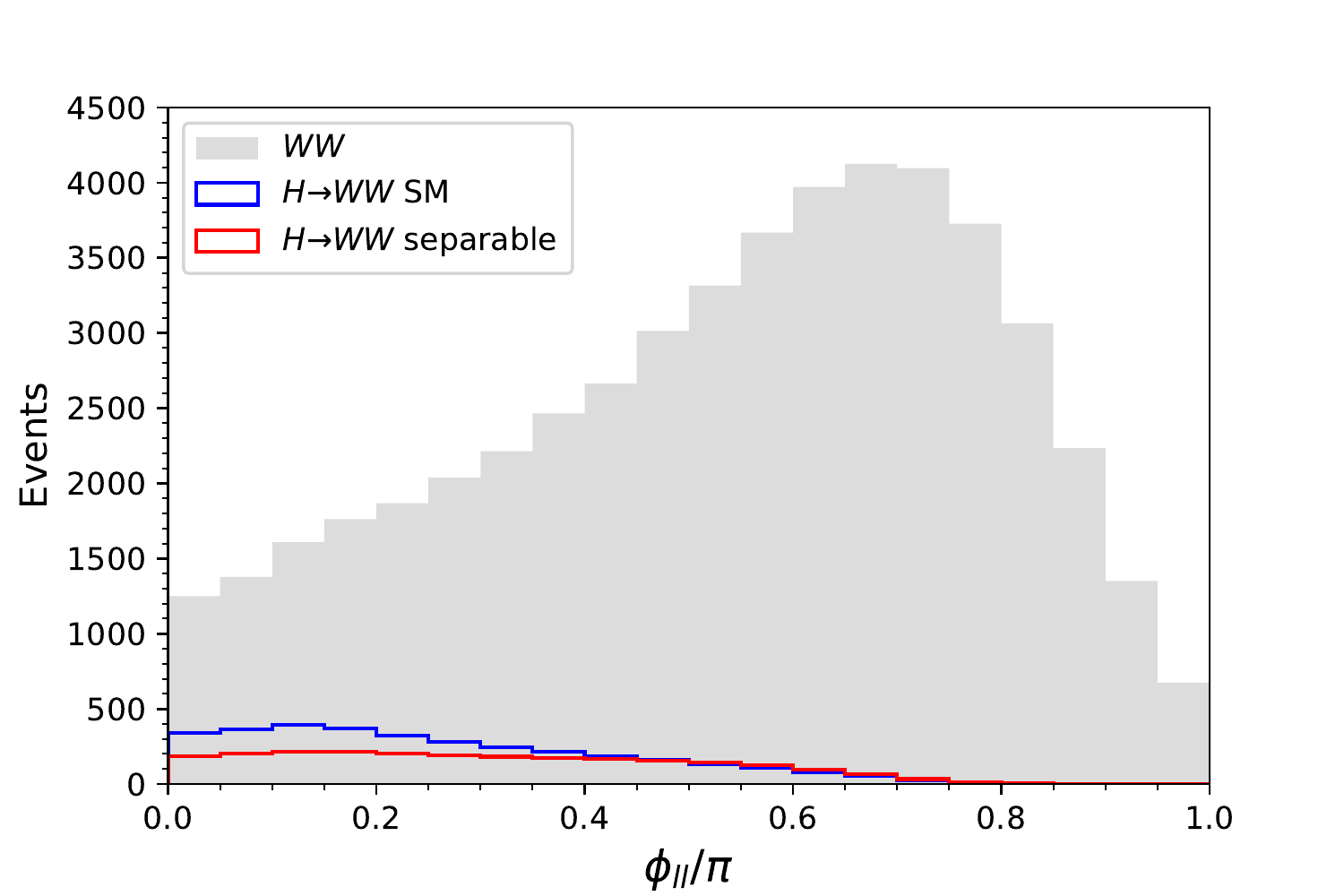} 
\end{tabular}
\caption{Dilepton observables: dilepton invariant mass (top) and azimuthal angle difference (bottom).}
\label{fig:distS}
\end{center}
\end{figure}

We present in Fig.~\ref{fig:distS} the kinematical distributions for $m_{\ell \ell}$ (top) and $\phi_{\ell \ell}$ (bottom) after simulation, for $H \to WW$ in the SM and the separable case, as well as for the $WW$ background. The luminosity is taken as $L = 138~\text{fb}^{-1}$. The differences in the shape of the $m_{\ell \ell}$ distribution observed at the parton level are maintained to a large extent. Moreover, the different angular distribution of the charged leptons leads to different event selection efficiencies (0.097 for the SM and 0.070 for the separable case) which also contribute to the discrimination between the two hypotheses. The striking differences in the shape of the $\phi_{\ell \ell}$ distribution that were observed at the parton level are washed out by the event selection, especially by the requirements on transverse masses. The $\eta_{\ell \ell}$ distribution turns out to be uninteresting because the SM and separable hypotheses are quite alike, and the signal concentrates near $\eta_{\ell \ell} \sim 0$ where the $WW$ background is also largest.

For the calculation of the expected statistical significance of the SM hypothesis over the separability we calculate the expected $\chi^2$ for the $WW+H$ (SM) versus the $WW + H$ (separable) hypotheses, using the ranges of $m_{\ell \ell}$ and $\phi_{\ell \ell}$ shown in Fig.~\ref{fig:distS}. This is a conservative approach since a narrower range would give larger deviations; on the other hand, the obtained estimation is more robust and less sensitive to the binning choice and possible mismeasurements of $m_{\ell \ell}$ and $\phi_{\ell \ell}$. 
For the $m_{\ell \ell}$ distribution we obtain $\chi^2 = 145$ for 14 degrees of freedom (d.o.f.) which amounts to a $7.1\sigma$ significance. (Selecting the range $10~\text{GeV} \leq m_{\ell \ell} \leq 40~\text{GeV}$ the statistical significance raises to $7.8\sigma$.) For the $\phi_{\ell \ell}$ distribution we obtain $\chi^2 = 76$ for 20 d.o.f., which amounts to $4\sigma$. 

The question that immediately arises is how systematic uncertainties may affect these estimates. In this regard,  the theoretical predictions for all processes are known at least to NNLO accuracy. The Higgs total production cross section can also be directly measured in other decay channels such as $H \to ZZ$. The normalisation for the background, from $WW$ and other processes, can be fixed by using different kinematical regions (see for example Ref.~\cite{CMS:2022uhn}), with scale factors that are close to unity.  Shape uncertainties have to be considered as well. 

\begin{figure}[t]
\begin{center}
\includegraphics[width=8cm,clip=]{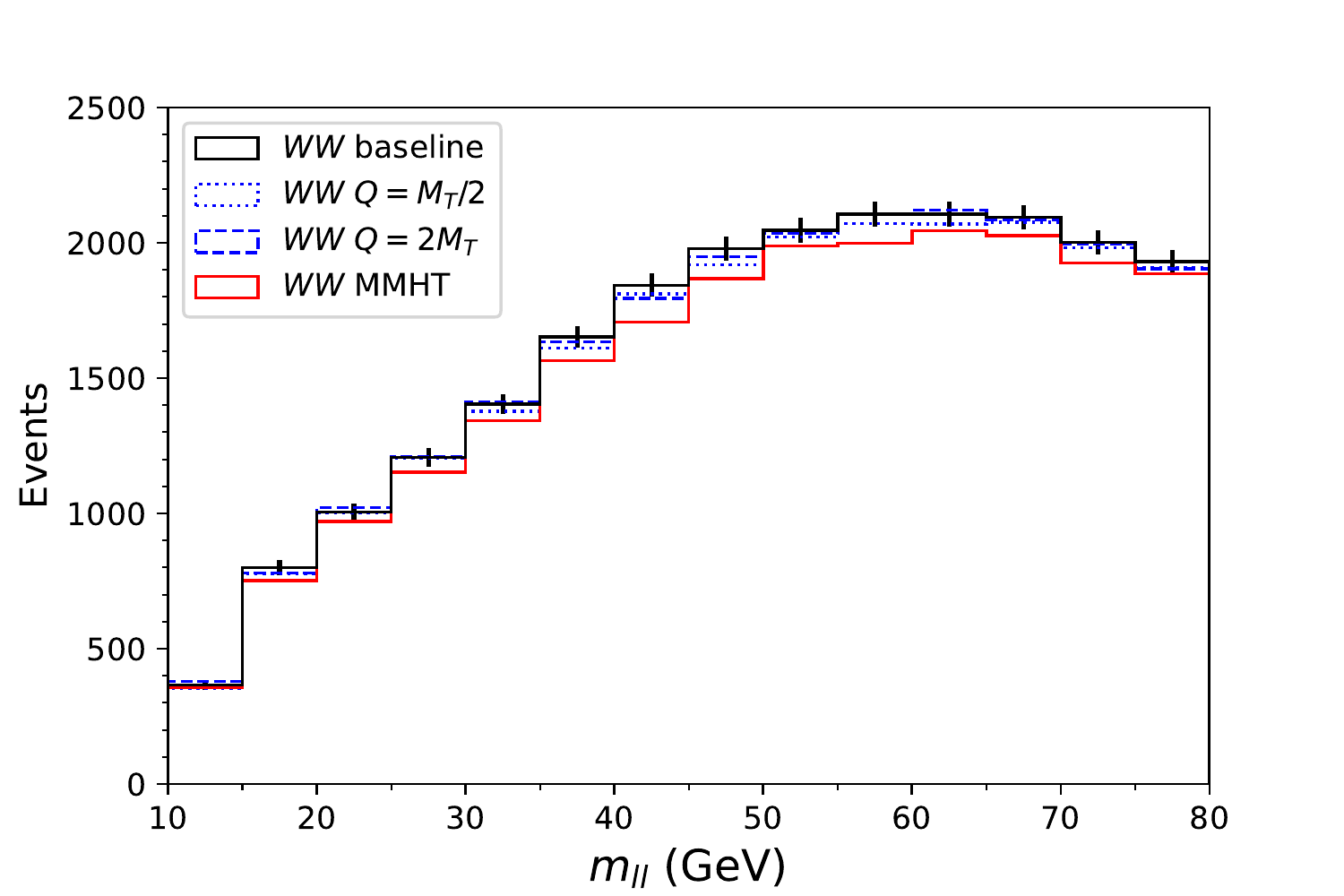} 
\caption{Dilepton invariant mass for the $WW$ background: baseline prediction and with different scales or PDFs. The vertical bars represent the statistical uncertainty for 138 fb$^{-1}$. }
\label{fig:WWsys}
\end{center}
\end{figure}

We have investigated the effect of theoretical shape uncertainties in the $m_{\ell \ell}$ distribution. The signal distribution is quite robust, as the dilepton invariant mass from the on-shell Higgs decay is determined by the decay kinematics. On the other hand, uncertainties in the $WW$ background may affect the signal extraction. We have investigated the uncertainty associated to:
\begin{itemize}
\item Changing the factorisation and renormalisation scale from the total transverse mass $M_T$ (default) to $M_T/2$ and $2 M_T$.
\item Replacing the baseline NNPDF 3.1~\cite{NNPDF:2017mvq} parton density functions (PDFs) by MMHT 2014~\cite{Harland-Lang:2014zoa}.
\end{itemize}
All the alternative $WW$ samples are generated with  $3 \times 10^6$ events.
We present in Fig.~\ref{fig:WWsys} the distribution for the $WW$ background in the relevant region $m_{\ell \ell} \leq 80$ GeV. The relative size of the samples has been normalised to the same cross section in the sideband $m_{\ell \ell} \in [80,150]$ GeV.\footnote{This procedure can be performed directly in data. We note that the statistical uncertainty associated to the sideband normalisation is small, below 0.8\% in the examples discussed.}

In the most relevant region of small $m_{\ell \ell}$ (where the SM and separable cases are better discriminated, see Fig.~\ref{fig:distS}) the theoretical uncertainties are small, of the order of the statistical uncertainty. Therefore, it is expected that theoretical uncertainties do not spoil the discrimination between the two hypotheses. We calculate the resulting $p$-value by using a Bayesian approach~\cite{Demortier:1099967}, assuming a flat prior for the different SM predictions. With the inclusion of the above discussed uncertainties, the $p$-value for the comparison between the SM and separable hypotheses slightly drops to $6.1$ standard deviations. 
An estimation of experimental systematic uncertainties can only be done with a full detector simulation and is beyond the scope of this work.

\section{Discussion}
\label{sec:5}

Generically, the test of quantum properties such and entanglement and violation of Bell inequalities requires the measurement of spin correlation observables. This, in turn, requires the reconstruction of rest frames and thus the full reconstruction of the relevant event kinematics. For example, the measurement of the $C_{L_1 M_1 L_2 M_2}$ coefficients in (\ref{ec:dist4D}) can be done by integration with spherical harmonics, which in turn requires knowledge of the angles $\theta_i$ and $\phi_i$ in the respective $V_i$ rest frames. For the decay $H \to ZZ \to 4\ell$~\cite{Aguilar-Saavedra:2022wam} this is not a problem, but for $H \to WW \to \ell^+ \nu \ell^- \nu$ the presence of the two neutrinos makes a unique reconstruction of the kinematics simply not possible. A probabilistic approach using a kinematical fit or a multivariate method remains to be explored.

Still, in the particular case of $H \to V_1 V_2$ decays there is a unique characterisation of the entanglement: as we have shown in section~\ref{sec:2}, the separability condition $C_{222-2} = 0$~\cite{Aguilar-Saavedra:2022wam} implies that only one of the three decay amplitudes, namely with both bosons longitudinally polarised, is non-zero. Thus, we can reformulate the entanglement condition as a binary test: SM versus longitudinal polarisation. And this binary test can be performed using laboratory-frame observables, as shown in section~\ref{sec:3}. For the specific case of the dilepton invariant mass, which is a quite robust variable already measured by the ATLAS and CMS Collaborations, the expected significance between the two hypotheses is of $6.1\sigma$ with a luminosity of 138 fb$^{-1}$. This figure includes statistical uncertainties, as well as an estimation of shape systematics from modelling. Therefore, the entanglement in $H \to WW$ could be established with the already collected Run 2 data. 

\section*{Acknowledgements}
I thank A. Bernal, J.A. Casas and J.M. Moreno for previous collaboration and many useful discussions.
This work is supported by the grants IFT Centro de Excelencia Severo Ochoa SEV-2016-0597, CEX2020-001007-S and PID2019-110058GB-C21 funded by MCIN/AEI/10.13039/501100011033 and by ERDF, and by FCT project CERN/FIS-PAR/0004/2019.

\bibliographystyle{utphys}
\bibliography{references}

\end{document}